\documentclass[prb,amsmath,amssymb,twocolumn,floatfix]{revtex4}
\usepackage{graphicx}
\usepackage{dcolumn}
\usepackage{bm}
\usepackage{subfigure}
\usepackage{amsmath}
\usepackage{hyperref}
\usepackage{times}
\usepackage{epsfig}
\usepackage{epstopdf}
\usepackage{graphicx} 

\hypersetup{
    bookmarks=true,         
    unicode=false,          
    pdftoolbar=true,        
    pdfmenubar=true,        
    pdffitwindow=true,      
    pdftitle={My title},    
    pdfauthor={Author},     
    pdfsubject={Subject},   
    pdfcreator={Creator},   
    pdfproducer={Producer}, 
    pdfkeywords={keywords}, 
    pdfnewwindow=true,      
    colorlinks=true,       
    linkcolor=red,          
    citecolor=blue,        
    filecolor=magenta,      
    urlcolor=cyan           
}
\DeclareMathAlphabet{\bi}{OML}{cmm}{b}{it}

\begin{document}
\def\ea{\textit{et al.}}
\def\bDel{{\vec{\mathbf{\Delta}}}}
\def\bk{{\bf k}}
\def\bz{{\bf z}}
\def\bG{\hat{{\bm G}}}
\def\bR{{\bf R}}
\def\br{{\bf r}}
\def\bj{{\bf j}}
\def\bx{\mathbf{\hat{x}}}
\def\by{\mathbf{\hat{y}}}
\def\bq{{\bf q}}
\def\bp{{\bf p}}
\def\bea{\begin{align}}
\def\eea{\end{align}}
\def\beq{\begin{equation}}
\def\eeq{\end{equation}}
\def\bdm{\begin{displaymath}}
\def\edm{\end{displaymath}}
\def\bJ{{\bf J}}
\def\bU{{\bf U}}
\def\Sr{\mathrm{Sr}_2\mathrm{RuO}_4}
\def\Up{\mathrm{UPt}_3}
\def\P{\hat{\Psi}}
\def\b0{\mathbf{0}}

\newcommand{\ua}{\uparrow}
\newcommand{\da}{\downarrow}
\newcommand{\ra}{\rightarrow}
\newcommand{\la}{\leftarrow}
\newcommand{\bs}{\boldsymbol}
\newcommand{\ep}{\epsilon}
\newcommand{\ld}{\lambda}

\title{Vanishing edge currents in non-$p$-wave topological chiral superconductors}
\author{Wen Huang$^1$, Edward Taylor$^1$ and Catherine Kallin$^{1,2}$}
\affiliation{$^1$Department of Physics and Astronomy, McMaster University, Hamilton, Ontario, L8S 4M1, Canada}
\affiliation{$^2$Canadian Institute for Advanced Research, Toronto, Ontario M5G 1Z8, Canada}
\date{Dec. 20, 2014}

\begin{abstract}
The edge currents of two dimensional topological chiral superconductors with nonzero Cooper pair angular momentum---e.g., chiral $p$-, $d$-, and $f$-wave superconductivity---are studied.  Bogoliubov-de Gennes and Ginzburg--Landau calculations are used to show that in the continuum limit, \emph{only} chiral $p$-wave states have a nonzero edge current.  Outside this limit, when lattice effects become important, edge currents in non-$p$-wave superconductors are comparatively smaller, but can be nonzero.  Using Ginzburg--Landau theory, a simple criterion is derived for when edge currents vanish for non-$p$-wave chiral superconductivity on a lattice.  The implications of our results for putative chiral superconductors such as $\Sr$ and $\Up$ are discussed. 
\end{abstract}

\maketitle

\section{Introduction}
Two-dimensional topological chiral superconductors break time-reversal symmetry by virtue of the fact that the Cooper pairs have nonzero orbital angular momentum.  For simple orbital eigenstates of the ($z$-component of the three-dimensional) angular momentum operator such as $p$-, $d$-, and $f$-wave states, the Cooper pairs each carry $m \hbar$ of angular momentum, with nonzero integer magnetic quantum numbers $m$. In a finite sample of such a superconductor (for convenience, in this paper we will not distinguish between chiral superconductors and neutral chiral superfluids such as $^3$He, using ``superconductor'' to describe both), this Cooper pair orbital angular momentum is expected to give rise to a spontaneous edge current and related to this, a nonzero \emph{total} angular momentum.  

For $p$-wave superconductors, both the edge current and total angular momentum have been studied extensively (see e.g., Refs.~\onlinecite{Ishikawa:77,Mermin:80,Kita:98,Stone:04,Sauls:11}), largely due to the fact the chiral $p$-wave $A$ phase of $^3$He is the only system which is known to be definitely chiral.  At the same time, the perovskite superconductor $\Sr$ is widely believed to be chiral $p$-wave~\cite{Mackenzie:03, Maeno:12,Kallin:12}, although magnetic fields consistent with the expected edge current have yet to be detected~\cite{Kirtley:07,Curran:14,Kallin:09}.  This last fact in particular has generated considerable interest in the question of what exactly is the relationship between topological chiral superconductivity and edge currents.  Although it can be strongly suppressed by disorder~\cite{Ashby:09,Sauls:11} as well as gap anisotropy and band effects~\cite{Huang:14}, the edge current and total angular momentum of a chiral $p$-wave superconductor are generically \emph{large}, the latter for instance being $L_z = N\hbar /2$~\cite{Stone:04,Stone:08} in the continuum limit for an ideal surface at $T=0$, where $N$ is the total number of fermions.  

In this paper, we generalize previous studies of the edge current in chiral $p$- and $d$-wave superconductors~\cite{Matsumoto:95,Rainer:98,Horovitz:03,Braunecker:05}.  In addition to being a problem of intrinsic theoretical interest, giving greater insight into the nature of the edge current in chiral $p$-wave superconductors for instance, this work will be relevant in the quest to find non-$p$-wave chiral superconductors such as the possibly chiral $f$-wave superconductor $\Up$~\cite{Joynt:02,Schemm:14}.   In contrast to the generically large edge current in chiral $p$-wave superconductivity, we find that the edge current in states with higher orbital Cooper pair angular momentum can vanish, depending on details of the lattice.  All our results are for unscreened currents.

Drawing on analytic semiclassical Bogoliubov-de Gennes (BdG) and Ginzburg--Landau (GL) calculations for continuum systems, we show that, amongst chiral pairing states that are eigenstates of the angular momentum operator, only chiral-$p$ superconductors have a nonzero edge current.  Our results extend to three-dimensional (3D) superconductors by considering eigenstates of the $z$-component $\hat{L}_z$ of the orbital angular momentum operator: only states with magnetic quantum number $m=1$ give rise to a nonzero edge current.  This means e.g., that the 3D $f$-wave state $k_z^2(k_x+ik_y)$ with $m=1$ has an edge current, but the $m=2$ state $k_z[(k^2_x-k^2_y) \pm 2ik_xk_y]$ does not.  The latter is the continuum analogue of a possible order parameter for $\Up$.

Turning to lattice models, numerical BdG and GL calculations are used to understand how these results carry over from the continuum.  Away from the continuum limit, the edge current  along axes of high symmetry can be nonzero even for non-$p$-wave chiral states, although for all cases studied, it is reduced as compared to that for chiral $p$-wave on a square lattice.  In some cases, such as chiral $f$-wave on a triangular lattice, we find that  the integrated current is extremely small.  In all cases where we find such a small integrated current, the local current oscillates over a small length scale comparable to the lattice spacing with an amplitude that decreases linearly with $\Delta_0/E_F$~\cite{Braunecker:05} and hence, vanishes in the weak-coupling limit.  A general condition for which the edge current vanishes consistent with our BdG results is derived within GL theory.

We start in Sec.~\ref{sec:continuum} by presenting our semiclassical analysis for systems in the continuum limit.  The implications of our results for the problem of the total angular momentum are discussed in Sec.~\ref{sec:Lztotal}.  There, a Chern--Simons-like~\cite{Volovik:88,Goryo:98,Stone:04,Huang:14} expression for the current is also discussed in connection with the possibility of a ``soft'' edge, where the density vanishes slowly as compared to the coherence length.  Apart from this section, and also a brief discussion given in Sec.~\ref{sec:GL}, we leave implicit that all our results are for a sharp edge, where the density vanishes over a distance on the order of the mean interparticle spacing $k^{-1}_F$.

Turning our focus to lattice models, in Sec.~\ref{sec:lattice}, results are given for numerical BdG calculations of the edge current  for chiral $p$-, $d$-, and $f$-wave order parameters in some representative lattice systems: $p_x+ip_y$ on a square lattice, $d_{x^2-y^2}+id_{xy}$ on square and triangular lattices, and $f_{x(x^2-3y^2)}+if_{y(3x^2-y^2)}$ on a triangular lattice.  In Sec.~\ref{sec:GL}, we reproduce our continuum as well as numerical lattice BdG results using GL theory.  A summary of our results is given in Sec.~\ref{sec:discussion} along with a discussion of their relevance for systems such as $\Sr$ and $\Up$, which have been proposed as candidate chiral superconductors.

\section{Edge current in the continuum limit of chiral superconductors}
\label{sec:continuum}

We begin by using semiclassical Bogoliubov-de Gennes calculations to understand properties of the edge current for an edge in two-dimensional continuum chiral superconductors.  For continuum systems, the Cooper pair eigenstates 
\begin{equation}
\Delta_{\bk} = \Delta_0 \left(\frac{k_x + i k_y}{k_F} \right)^m \equiv \Delta_0(k/k_F)^me^{i m\theta} \,\,,\,\,\,m=1,2,...
\label{OPcont}
\end{equation}
of the 2D angular momentum operator are characterized by the magnetic quantum number $m$.  $\theta$ is defined such that $\bk = k[\cos\theta,\sin\theta]$.  Not only does the magnetic quantum number give the angular momentum $m\hbar$ per Cooper pair, it also is equal to the Chern number (or skyrmion number of the BdG Hamiltonian)~\cite{Volovik:99},
\beq
m=C \equiv \frac{1}{4\pi}\int d^2k\; \hat{h} \cdot \left( \partial_{k_x} \hat{h} \times \partial_{k_y} \hat{h}\right),\label{Chern}\eeq
which counts the number of zero-energy edge modes.  
Here $\vec h =\left\{\mathrm{Re}[\Delta_{\bk}],-\mathrm{Im}[\Delta_{\bk}],\xi_{\bk}\right\}$ and $\hat h = \vec h/|\vec h|$, with $\xi_{\bk} \equiv \epsilon(\bk)-\mu$ the single-particle dispersion. 

\begin{figure}
\includegraphics[width=0.85\linewidth]{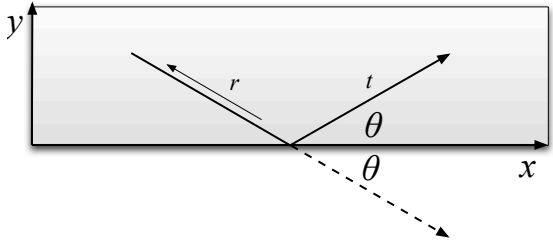}
\caption{Specular ($t$) and Andreev ($r$) reflection of a quasiparticle off an ideal edge at $y=0$. Adapted from Ref.~\onlinecite{Stone:04}.}
\label{fig:A1}
\end{figure}

The Bogoliubov-de Gennes (BdG) equation for the order parameter (\ref{OPcont}) is
\beq \left[\begin{array}{cc}h_0 & \Delta_0\left(\frac{k}{k_F}\right)^m e^{i m\theta} \\ \Delta_0\left(\frac{k}{k_F}\right)^me^{-im\theta} & -h^*_0\end{array}\right]\left[\begin{array}{c}u\\ v\end{array}\right] = E\left[\begin{array}{c}u\\ v\end{array}\right],\label{BdG}\eeq
where $h_0 \equiv -\tfrac{\hbar^2}{2m^*}\nabla^2-\mu$ and we have used $m^{*}$ to denote the fermion mass to avoid confusion with the magnetic quantum number.  We seek solutions of (\ref{BdG}) for the situation where there is an edge parallel to the $x$-axis, at $y=0$.  This edge is implemented using the boundary condition $u(y=0)=v(y=0)=0$. 

A spontaneous current arises at an edge due to both current-carrying Andreev-scattered edge states as well as the reflection of continuum states~\cite{Stone:04}.  The corresponding solutions 
\beq \P = \sum_{\sigma=\pm }\sigma \left[\begin{array}{c} a_{\sigma}(y)\\ b_{\sigma}(y)\end{array}\right]e^{ik_Fx\cos\theta + i\sigma k_Fy\sin\theta}\label{BdGsolns}\eeq
of the BdG equations are thus completely parameterized by the incident angle $\theta$; see Fig.~\ref{fig:A1}.  In (\ref{BdGsolns}), the $\sigma=\pm $ components of the solution represent the transmitted (specular reflection) and reflected (Andreev reflection) solutions, respectively.  Note that for our chosen geometry, this angle is the same as the one that enters the order parameter (\ref{OPcont}).  The minus sign ($\sigma=-1$) attached to the reflected solution means that the vanishing of the wavefunction at the edge becomes  $\P_-(0) = \P_+(0)$, where $\P^{\dagger}_{\sigma}\equiv [a_{\sigma},b_{\sigma}]$.  The current density \emph{per spin component} corresponding to this solution is thus
\begin{align} j_x(y>0) &= \frac{\hbar}{4m^{*} i }\left[\P^{\dagger}\partial_x\P - (\partial_x\P^{\dagger})\P\right]\nonumber\\&=\frac{\hbar k_F\cos\theta}{2m^*}\sum_{\sigma=\pm}\P^{\dagger}_{\sigma}\P_{\sigma}.\label{jdens}
\end{align}
As noted in Ref.~\onlinecite{Stone:04}, the seemingly extra factor of $1/2$ in this expression is needed to compensate the double-counting in the particle-hole basis spanned by $\P$.  

To solve the BdG equations, (\ref{BdG}) and (\ref{BdGsolns}), we adopt the elegant approach used by Stone and Roy~\cite{Stone:04} to solve the $m=1$ problem and map these equations onto the one-dimensional ``twisted mass'' Dirac problem. The density $\sum_{\sigma}\P^{\dagger}_{\sigma}\P_{\sigma}$ of quasiparticle states receives contributions from the bound edge state as well as the ``charge''  $Q_m(\theta)$ arising from the phase-shifted bulk continuum states that accumulates at the edge.  Each bound state has unit normalization and thus its contribution to the integrated current is obtained by integrating (\ref{jdens}) over the values of $\theta$ for which the edge mode spectrum is negative (i.e., occupied):
\beq J_{\mathrm{edge}} = \int_\text{occupied} \frac{k_F \sin \theta d\theta}{2\pi} \left(\frac{\hbar k_F \cos\theta}{2m^*}\right) .\label{Jedge1}\eeq
The contribution to the current from bulk continuum states is similarly 
\beq
 J_{\mathrm{bulk}} =\int_0^{\pi} \frac{k_F \sin \theta d\theta}{2\pi}Q_m(\theta) \left(\frac{\hbar k_F\cos\theta}{2m^{*}} \right).
\label{Jbulk1}
\eeq

In Appendix~\ref{sec:semiclassical} we use the solutions of the twisted-mass Dirac problem to show that the edge mode spectrum and accumulated charge are given by piecewise functions 
\beq E^{(0)}=  (-1)^j \Delta_0 \cos (m\theta)\;\;\mathrm{for}\;\frac{(j-1)\pi}{m} \leq \theta<\frac{j\pi}{m}  \label{dispersion}\eeq
and
\beq Q_m(\theta) =  \frac{m\theta}{\pi} - j\;\;\mathrm{for}\;\frac{(j-1)\pi}{m} \leq \theta<\frac{j\pi}{m},\label{Q}\eeq
with $j=1...m$.  The edge mode dispersion means that the occupied edge states correspond to incident angles $\theta\in [0,\pi/2m],~[\pi/m,3\pi/2m]$,..., $[(m-1)\pi/m,(m-1/2)\pi/m]$, and (\ref{Jedge1}) becomes
\beq J_{\mathrm{edge}}=  \frac{\hbar k_F^2}{16\pi m^{*}} \sum_{j=1}^{m} \left[\cos\frac{(2j-2)\pi}{m}-\cos\frac{(2j-1)\pi}{m} \right].\label{Jedge2}\eeq
Using (\ref{Q}) in (\ref{Jbulk1}), the bulk state contribution to the current is
\begin{align} J_{\mathrm{bulk}} =  -\frac{\hbar k_F^2}{4\pi m^\ast} \sum_{j=1}^{m} \Bigg[& \frac{m}{8\pi} \left(\sin\frac{(2j-2)\pi}{m} - \sin \frac{2j\pi}{m} \right)\nonumber\\&+\frac{1}{4}\cos\frac{(2j -2)\pi}{m} \Bigg].\label{Jbulk2}\end{align}

For chiral $p$-wave ($m=1$),  the bulk contribution is half in magnitude as the current carried by the chiral edge states, and flows in the opposite direction: $J_{\mathrm{edge}} = \hbar k_F^2/(8\pi m^{*})$ and $J_{\mathrm{bulk}} = -\hbar k^2_F/(16\pi m^{*})$~\cite{Stone:04}.  The total edge current per spin component can thus be written as $J =n\hbar/4m^*$, where $n=k_F^2/4\pi$ is the number density per spin component. This value is consistent with numerical BdG calculations in the continuum limit of lattice models~\cite{Huang:14} (for simple lattice models at least, iterating BdG to full self-consistency has negligible impact on our results).  It is also the edge current needed to produce a macroscopic angular momentum $N\hbar/2$ for $N$ fermions in a disc~\cite{Stone:04} (see below). 

On the other hand, the edge state and continuum state contributions (\ref{Jedge2}) and (\ref{Jbulk2}) vanish independently \emph{for all} $m>1$, a fact that can be proved by induction. Thus the total edge current is identically zero for any chiral superconductor with Cooper pair angular momentum $>\hbar$.  Note that although multiple chiral edge branches with the same chirality exist for $m>1$, the contributions to the current exactly cancel among those chiral branches. In the continuum at least, $p$-wave is special\cite{Huang:14}! As noted in the Introduction, this result extends to 3D superconductors by considering eigenstates of the $z$-component $\hat{L}_z$ of the orbital angular momentum operator: only states with magnetic quantum number $m=1$ give rise to a nonzero edge current.

\section{Total angular momentum}
\label{sec:Lztotal}

Before discussing how the continuum limit results carry over to lattice models of chiral superconductivity, we briefly touch on a problem of some historic interest, namely the angular momentum carried by a disc of a neutral chiral superfluid\cite{footnote}.  The fact that the edge current vanishes for $m>1$ Cooper pair states means that a superfluid of $N$ fermions arising from these states will not have a macroscopic total angular momentum\beq L_z = \frac{N\hbar m}{2}\label{Lmexpected}\eeq
Such a macroscopic angular momentum would arise if there is a local current density~\cite{Stone:04,Sauls:11} $j(x)\sim N m v_F\Delta_0 \exp(-x/\xi_0)$ confined within a coherence length of the edge at weak-coupling.  It is moreover the expected result in the strong-coupling ``BEC limit''~\cite{Leggett:80,Mermin:80}, where the number of Cooper pairs (i.e., the condensate occupation) asymptotes to $N/2$.  For $p$-wave pairing, the edge current indeed gives rise to a total angular momentum given by (\ref{Lmexpected}) for both an ideal sharp edge~\cite{Stone:04,Sauls:11} as well as a soft one~\cite{Stone:08}.  For higher-angular momentum pairing, however, our BdG results suggest that (\ref{Lmexpected}) is not true in general.    

We define the total angular momentum of a disc of radius $R$ as
\beq L_z = \int_{r\leq R} d\br m^{*}(\br\times \bj)_z.\label{Lzdef}\eeq
Recall that $m^{*}$ is the fermion mass.
A nonzero local current $\bj(\br)$ only arises if the density or order parameters components vary in space.  Thus, for a disc having a sharp edge, wherein the density vanishes over an atomic scale at the edge, the only current is the edge current we have discussed in previous sections.  For higher-angular momentum Cooper pair states with $m>1$, the total angular momentum is zero.  

At the same time, if the edge is softened, such that the density vanishes over a length scale much longer than the BCS coherence length, the local edge current per spin component is given by~\cite{Volovik:88,Goryo:98,Stone:04,Huang:14}
\beq \bj(\br) = -\frac{\hbar C}{8\pi}(\hat{\bs z}\times\nabla)A_0(\br).\label{jCS}\eeq
Here $A_0(\br)$ is an external potential that gives rise to the slow density variation and $C$ is the Chern number (\ref{Chern}) which, as noted earlier, is equal to the magnetic quantum number $m$ in continuum systems for Cooper pair states that are eigenstates of the angular momentum. We have confirmed using numerical BdG (not shown) that the current is restored as the edge is softened, in agreement with the lattice discretized form of (\ref{jCS}), with $\partial_x A_0(x) \rightarrow A_0(x_{i+1})-A_0(x_i)$. Some discussion of the origin of this ``Chern--Simons-like'' contribution is given in Sec.~\ref{sec:GL}.

Using (\ref{jCS}) in (\ref{Lzdef}), for a rotationally-invariant potential $A_0(\br) = A_0(r)$, and using the equilibrium condition $\partial_r A_0(r) = (\partial\mu/\partial n)\partial_r n(r)$ with $\mu = 2\pi n/m^{*}$, the total angular momentum is
\beq L_z = -\frac{\hbar C m^{*}}{4} \int^{R}_0 dr r^2 \left(\partial \mu/\partial n\right)\partial_r n(r)=\frac{N \hbar C}{2},\label{LzC}\eeq
where $N = 2\pi \int^{R}_0 dr r n(r)$.  Thus, equating the Chern number with the magnetic quantum number $m$, when the density varies slowly, one recovers (\ref{Lmexpected}) for all cases with nonzero Cooper pair angular momentum.  It is only when the density varies sharply that the total angular momentum vanishes  for all states except $p$-wave.  

We note in passing that (\ref{jCS}) is equivalent to the ``intrinsic pair angular momentum'' identified by Mermin and Muzikar, arising from the orbital angular momentum of the Cooper pairs.  It indeed conspires to produce the expected macroscopic angular momentum (\ref{Lmexpected}) but only in general when the density varies slowly as compared to the BCS coherence length $\xi_0$.   Such a situation can arise, for instance, in an ultra-cold atomic gas chiral superfluid confined in harmonic traps~\cite{Stone:08}.

\section{Edge current for lattice models}
\label{sec:lattice}

We now turn to the question of whether our central continuum-limit result--the vanishing of the edge current in non-$p$-wave chiral superconductors--survives outside of this limit.  Some indication of the answer can be found in the literature, which has largely focussed on the possibility of chiral $d$-wave superconductivity in the cuprates~\cite{Matsumoto:95,Rainer:98,Horovitz:03} but also, more recently, chiral $d$-wave order in graphene~\cite{Jiang:08,Nandkishore:12,Black:12} and other materials~\cite{Braunecker:05,Kiesel:13,Fischer:14}.  A small (but nonzero) edge current along [11] surface was reported in Ref.~\onlinecite{Rainer:98} for chiral $d_{x^2-y^2}+id_{xy}$ superconductivity on a square lattice.  It is unclear, however, whether the calculation reported there allowed for the possibility that $d+is$ order (expected to produce a nonzero edge current~\cite{Rainer:98,Horovitz:03}) develops near the surface. In lattices with hexagonal symmetry, away from the continuum limit, Ref.~\onlinecite{Braunecker:05} finds a finite but small local current.  Nonzero edge currents are also found for chiral $d$-wave superconductivity on a honeycomb lattice~\cite{Black:12}.

Here we expand on these results, presenting numerical BdG calculations of the unscreened edge current in a few representative one-band models: chiral $p$- and $d$-wave on a square lattice, as well as chiral $f$- and $d$-wave on a triangular lattice.  The last has been proposed as a possible superconducting state in $\mathrm{Na}_{x}\mathrm{CoO}_2\cdot y\mathrm{H}_2\mathrm{O}$~\cite{Kiesel:13} and SrPtAs~\cite{Fischer:14}.  In contrast to $p$-wave pairing which has a large edge current along the axes of a square lattice, we find that the integrated edge current along the same axes is very small for $d_{x^2-y^2}+id_{xy}$ order, consistent with previous work~\cite{Horovitz:03}. The edge current is substantial for this state when placed on a triangular lattice, however. Considering chiral $f$-wave pairing on a triangular lattice, we find a very small integrated current.  In all cases where we find such a small integrated current, the local current varies rapidly over a scale $\sim k^{-1}_F$ with amplitude decaying linearly with $\Delta_0/E_F$, similar to that in Ref.~\onlinecite{Braunecker:05}.    We thus take our results to be indicative of a vanishing edge current in the weak-coupling limit of these cases.

\begin{figure}
\begin{center}
\includegraphics[width=0.85\linewidth]{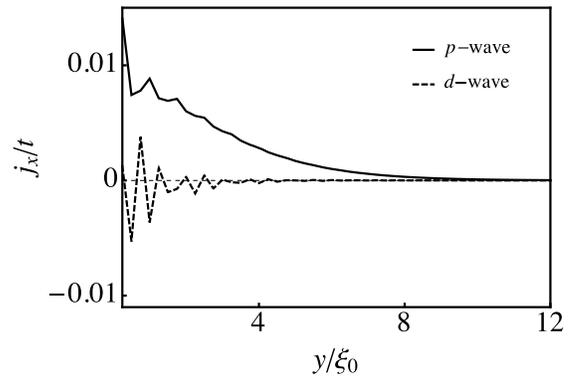}
\caption{Spatial dependence of the local edge current $j_x(y)$ for chiral $p$- and $d$-wave order parameters on a square lattice with hopping $t$.  The edge is at $y=0$ and the local currents extend over several coherence lengths $\xi_0 \equiv t/\Delta_0\sim 5$ (in units of the lattice spacing).  Calculations are done using $\mu=-t$ in conjunction with the order parameters described in the text for a strip of width 300 lattice sites along $y$ and with periodic boundary conditions along $x$.}
\label{fig:D1}
\end{center}
\end{figure}

\begin{figure}
\begin{center}
\includegraphics[width=0.85\linewidth]{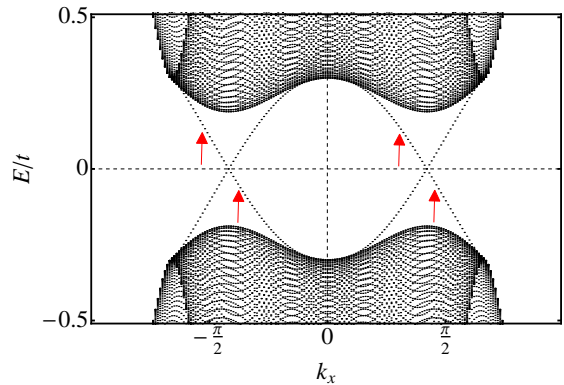}
\caption{Low energy dispersion of a one-band chiral $d$-wave model on a square lattice calculated using the same parameters used in Fig.~\ref{fig:D1}.  The arrows point to the chiral edge modes belonging to the same edge.
}
\label{Dspectrum}
\end{center}
\end{figure}

Our BdG calculations are carried out in the standard way  (see e.g., Ref.~\onlinecite{Lederer:14} for details) using a strip geometry, with edges at $y=0$ and $y=300$ (in units where the lattice spacing is 1), and periodic boundary conditions imposed along $x$.   Iterations are carried out to self-consistency.  Although sub-dominant orders can often be induced at the surface, we ignore these for simplicity.  For chiral $p_x+ip_y$ and $d_{x^2-y^2}+id_{xy}$ pairing on a square lattice, we use $\Delta_{\bk} = \Delta_0(\sin k_x+i\sin k_y)$ and $\Delta_{\bk} = \Delta_{01} (\cos k_x - \cos k_y) + i\Delta_{02} \sin k_x \sin k_y$, respectively.  These are allowed by the underlying tetragonal point group ($D_{4h}$) symmetry of the lattice; they reduce to $(k_x+ik_y)/k_F$ and $(k_x+ik_y)^2/k^2_F$ in the continuum limit. Note the two $d$-wave components are in general nondegenerate on a square lattice and $\Delta_{01}\neq \Delta_{02}$.  Using the same interaction strength for both channels, however, we find the $d_{xy}$ component to be too small to reliably carry out calculations.  To avoid this difficulty, we tune the interactions to give $\Delta_{01}\simeq \Delta_{02}$.  Changing these values does not affect our conclusion in cases where the edge current vanishes, however. In addition, the numerical calculations we present are for systems with one electron-like Fermi surface around the $\Gamma$ point. However, we have also done calculations for other scenarios and the discussion and conclusions which follow apply equally well to the general cases.

\begin{figure}
\centering
\begin{minipage}[r]{0.23\textwidth}
\includegraphics[width=1.0\linewidth]{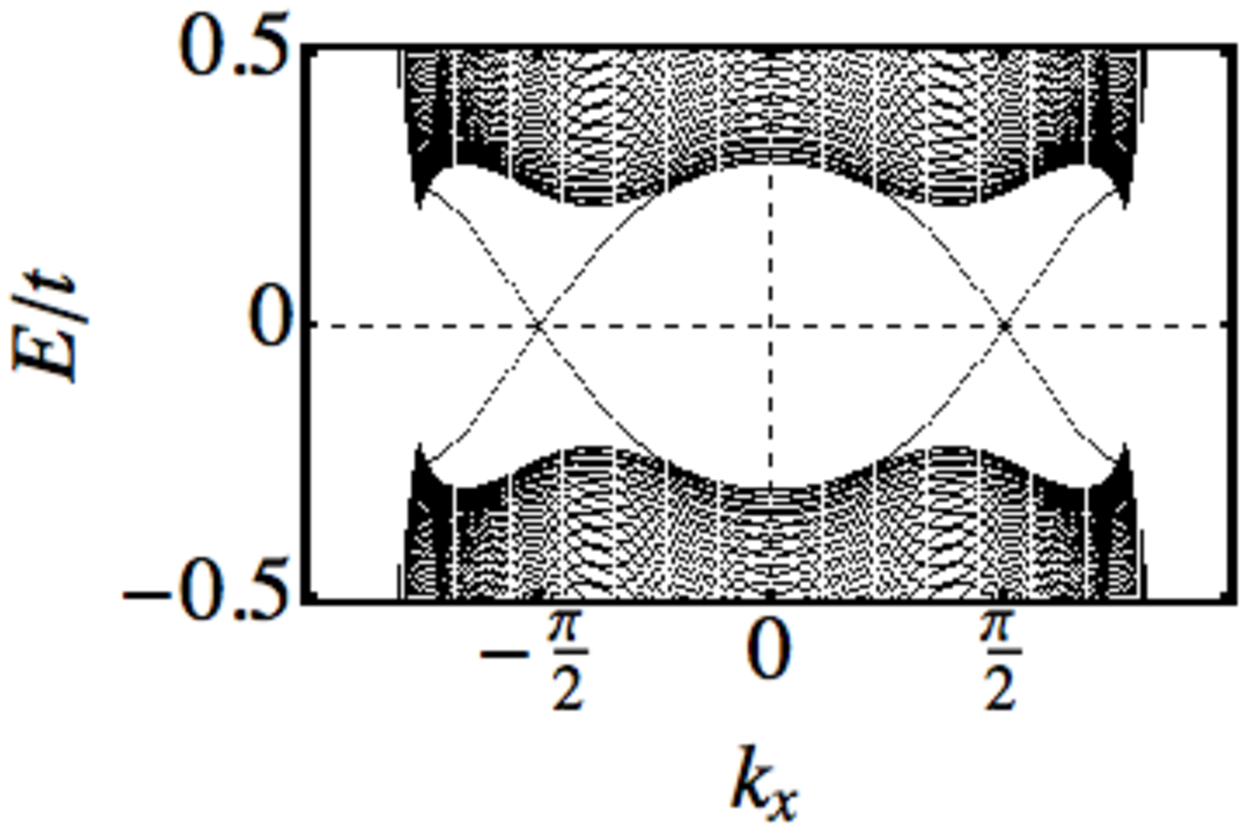}
\end{minipage}
\begin{minipage}[r]{0.21\textwidth}
\includegraphics[width=1.0\linewidth]{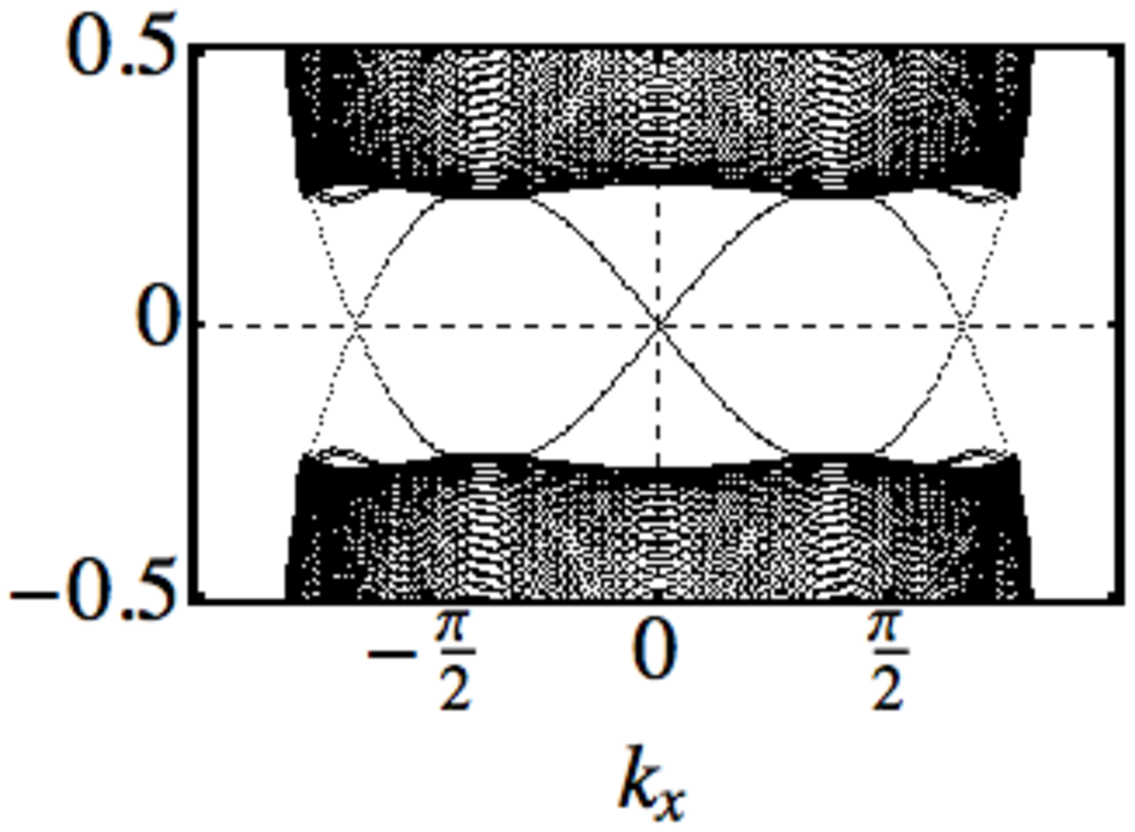}
\end{minipage}
\caption{ Edge dispersion of the chiral $d$- and $f$-wave models on a triangular lattice with the same parameters used in Fig.~\ref{fig:PDFcurr}.} 
\label{fig:FDdis}
\end{figure}

The local currents near the edge at $y=0$ for these two models are shown in Fig.~\ref{fig:D1}.  The local current for chiral $d$-wave oscillates with an amplitude that decays linearly with $\Delta_0$~\cite{Braunecker:05}.  In units of the lattice hopping $t$, the integrated current shown in Fig.~\ref{fig:D1} is $J \simeq 0.006t$, as compared to $J \simeq 0.12t$ for $p$-wave, and we expect that in the $\Delta_0\ll t$ limit, the integrated current vanishes for chiral $d$-wave on a square lattice.  This is true despite the fact that there are two chiral zero-energy (Majorana) bound state modes present on each edge; see Fig.~\ref{Dspectrum}.  In fact, for the contribution to the edge current from the chiral edge modes, it is precisely because there is more than one edge state that the contribution vanishes as a result of cancelling contributions.  As much is evident from the continuum-limiting expressions (\ref{dispersion}) and (\ref{Jedge2}) [we note that the former well-describes the in-gap dispersion shown in Fig.~\ref{Dspectrum} and also the spectra shown in Fig.~\ref{fig:FDdis} for $d$- and $f$-wave pairing on a triangular lattice].

\begin{figure}
\includegraphics[width=0.85\linewidth]{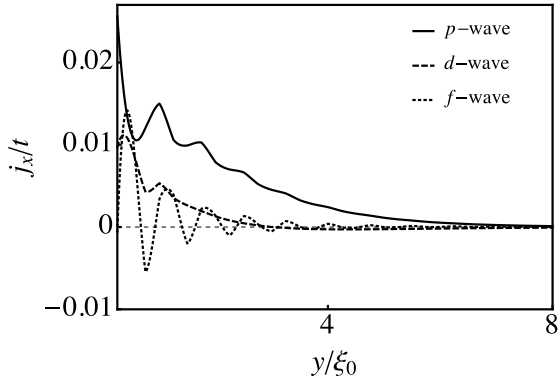}
\caption{Spatial dependence of the local edge current $j_x(y)$ for chiral $p$-, $d$- and $f$-wave order parameters on a triangular lattice with hopping $t$.  Calculations are done using $\mu=0$ and $\Delta_0 \approx 0.2t$ ($\xi_0 \equiv t/\Delta_0 \approx 5$) in conjunction with the order parameters described in the text for a strip with the same size as that used for the square lattice calculations. }
\label{fig:PDFcurr}
\end{figure}

For the triangular lattice, the chiral $d$-wave order takes the form of $\Delta_{\bk} = \Delta_0 \left[\cos k_x-\cos(\sqrt{3}k_y/2)\cos(k_x/2)\right] + i \Delta_0\sqrt{3}\sin \left(\sqrt{3}k_y/2\right) \sin(k_x/2)$, which also reduces to $(k_x+ik_y)^2$ in the continuum limit.  A chiral $f$-wave state of the form $\Delta_{\bk} = \Delta_{01}[\sin(2k_x) - 2 \cos (\sqrt{3}k_y) \sin k_x] 
+ i\Delta_{02}[ 2\sin (\frac{\sqrt{3}}{2}k_y)\cos (\frac{3}{2}k_x) -\sin(\sqrt{3}k_y)]$ can be realized on a triangular lattice with second and third neighbour odd-parity pairing. This gap function reduces to $(k_x+ik_y)^3$ in the continuum limit where the two components become degenerate.  Outside the continuum limit, the two order parameter components are not in general degenerate and $\Delta_{01}\neq \Delta_{02}$.  As with $d$-wave on a square lattice, we tune the interactions such that $\Delta_{01}\simeq \Delta_{02}$.  In Fig.~\ref{fig:PDFcurr} we plot the edge currents of the chiral $d$- and $f$-wave models on a triangular lattice with an edge along one side of the triangles.  For comparison, we also plot the edge current of a chiral $p$-wave superconductor, with $\Delta_{\bk}=\Delta_0[\sin(\sqrt{3}k_y/2)\cos (k_x/2)+\tfrac{i}{\sqrt{3}}(\sin k_x + \cos(\sqrt{3}k_y/2)\sin(k_x/2))]$.  As with $d$-wave, the two order parameter components are degenerate on a triangular lattice.  While the $p$- and $d$-wave models do not yield vanishing edge currents, the local  edge current for the chiral $f$-wave state oscillates rapidly about zero, integrating to a small value, $J \simeq 0.017t$, much smaller than the corresponding value ($J\simeq 0.15t$) for $p$-wave and about half the size of the value ($J\simeq 0.036t$) for $d$-wave.  As with our chiral $d$-wave results on a square lattice, we interpret this result as meaning that the edge current vanishes in the weak-coupling limit for chiral $f$-wave on a triangular lattice. 

Even though the edge current for chiral $d$-wave on a triangular lattice is nonzero, it is smaller than that for $p$-wave.  Moreover, consistent with our semiclassical analysis and also  Ref.~\onlinecite{Braunecker:05}, it vanishes in the weak-coupling, continuum limit, as $\mu$ approaches the bottom of the band.  

\section{Ginzburg-Landau theory}
\label{sec:GL}
We now seek insight into our BdG results from Ginzburg--Landau (GL) theory.  The current arises from gradient terms in the GL free energy density.   For a system with a two (complex) component order parameters $\psi_1$ and $\psi_2$, ignoring the possibility of an external potential, $A_0(\br)=0$, the terms responsible for the current are~\cite{Sigrist:91}
\begin{align}
f_{\mathrm{GL}}= k_3 (  \partial_x \psi^\ast_1 \partial_y \psi_2 + \mathrm{c.c.}) +  k_4( \partial_y \psi^\ast_1 \partial_x \psi_2  + \mathrm{c.c.})+\cdots
\label{eq:GL1}
\end{align}
where the ellipsis denotes higher-order terms.  Making contact with our microscopic results, the complex order parameter is
\beq [\psi_1(\br),\psi_2(\br)]\equiv [\Delta_{01}(\br),i\Delta_{02}(\br)]\exp[i\theta(\br)],\label{GLOP}\eeq
where $\theta(\br)$ is the $U(1)$ phase and $\Delta_{01}(\br)$ and $\Delta_{02}(\br)$ are the purely real, spatially varying amplitudes, reducing to the bulk values $\Delta_{01}$ and $\Delta_{02}$ away from an external potential and far from the edge.  

We emphasize that even though the notation of (\ref{eq:GL1}) is usually reserved for systems with tetragonal symmetry (see e.g., Table VII in Ref.~\onlinecite{Sigrist:91}), one can always construct an expression of the form given by (\ref{eq:GL1}) and it is valid for systems with arbitrary lattice symmetry.  Adopting the notation in Ref.~\onlinecite{Sigrist:91} for instance, our $k_3$ and $k_4$ are equal to $K_3$ and $K_4$ for a tetragonal lattice; for a hexagonal lattice, terms of the form (\ref{eq:GL1}) also arise however one instead has $k_3 = K_1 - K_3$ and $k_4 = -K_2+K_3$.  Moreover, to leading order in the gap amplitudes $\Delta_0$, $k_3$ and $k_4$ are equal~\cite{Furusaki:01}.  

Using (\ref{GLOP}), the $\mu$-component of the current is  (where it appears as a Cartesian index, $\mu,\nu=1,2$ denote the $x,y$ axes) is
\beq j_{\mu} = \frac{\partial f_{\mathrm{GL}}}{\partial (\partial_{\mu}\theta)} = k_3\epsilon_{\mu\nu}(\Delta_{0\mu}\partial_{\nu}\Delta_{0\nu}- \Delta_{0\nu}\partial_{\nu}\Delta_{0\mu}),\label{jGL}\eeq
where $\epsilon_{\mu\nu}$ is the 2D Levi--Civita symbol.  
Hence, a vanishing edge current along one of the crystalline axes is associated with the vanishing of the $k_3$ GL coefficient.  

As in Ref.~\onlinecite{Furusaki:01}, the GL expression (\ref{jGL}) serves as an alternative and more phenomenological description of the BdG current.  Although (\ref{jGL}) is only rigorously valid close to $T_c$ and does not give the exact current at low temperatures, it has been well established that GL theory provides a reliable qualitative description of the current in BdG calculations~\cite{Furusaki:01,Lederer:14}, and this is also confirmed here.

The gradient terms (\ref{eq:GL1}) in the GL free energy density lead to the following microscopic expression for $k_3$:
\beq k_3 = k_4=\left.\frac{\partial^2}{\partial q_x q_y}\Gamma^{-1}_{12}(\bq,0)\right|_{T=T_c},\label{k3def}\eeq
where
\beq \Gamma^{-1}_{\alpha\beta}(\bq,0) = -\sum_{\bk}\frac{h_{\alpha}(\bk)h_{\beta}(\bk)(1-f_{\bk}-f_{\bk-\bq})}{\xi_{\bk}+\xi_{\bk-\bq}} + \frac{\delta_{\alpha\beta}}{g}\label{ppvertex}\eeq
is the inverse of the static particle-particle vertex function in the $\alpha$-$\beta$ Cooper pair channel.  
$h_{\alpha}(\bk)$ are the dimensionless form factors that arise in the order parameter components, $\vec{\Delta}_{\bk} = [\Delta_{01}h_1(\bk),i\Delta_{02}h_2(\bk)]$, and also the attractive interaction $V_{\alpha}(\bk,\bk') = -g h_{\alpha}(\bk)h_{\alpha}(\bk')$ in the relevant channel; $f_{\bk} = [\exp(\beta \xi_{\bk})+1]^{-1}$ is the Fermi occupation.  Applying (\ref{k3def}) to (\ref{ppvertex}) gives
\begin{align}
k_3=\sum_{\bk}\frac{h_{1}(\bk)h_{2}(\bk)}{8\xi^3_{\bk}}\Bigg\{&v_x v_y \left[\beta_c XY\xi^2_{\bk} +Y\xi_{\bk}-2X\right]\nonumber\\&+(\partial_{k_x}v_y)\left[2X\xi_{\bk}-Y\xi^2_{\bk}\right]\Bigg\}.\label{k3}
\end{align}
Here, $v_i\equiv \partial_{k_i}\xi_{\bk}$, $X\equiv\tanh(\beta_c\xi_{\bk}/2)$, and $Y\equiv \beta_c\mathrm{sech}^2(\beta_c \xi_{\bk}/2)$, with $\beta_c \equiv T^{-1}_c$.  

Of all eigenstates of the $z$-component of the angular momentum operator $\hat{L}_z$, (\ref{k3}) confirms that chiral $p$-wave, with eigenvalue $m=1$, is special.  Using the continuum-limit form (\ref{OPcont}), $h_1(\bk) = \cos m\theta$ and $h_2 = \sin m\theta$.   Using $v_x\propto k\cos\theta$ and $v_y\propto k\sin\theta$, $k_3$ can be written as
\beq k_3 =  I(\mu,T_c)\int^{2\pi}_0 d\theta \sin \theta \cos \theta \sin m\theta \cos m\theta,\label{k3cont}\eeq
where $I(\mu,T)$ is an integral over the radial part of $\bk$.  This shows explicitly that $k_3$ vanishes in the continuum limit for all $m$ except 1~\cite{note}, in agreement with our semiclassical BdG results in Sec.~\ref{sec:semiclassical}, showing that the edge current vanishes for all $m\neq 1$.  

Moving away from the continuum limit, (\ref{k3}) remains valid for lattice systems using the appropriate forms for $h_1,h_2$, and $\xi_{\bk}$.  The condition for $k_3$ to vanish becomes more complicated than the continuum result (\ref{k3cont}), however.   More generally, noting that the integrand in (\ref{k3}) is strongly peaked about the Fermi surface and that the second line vanishes under particle-hole symmetry, GL theory predicts that the edge along a crystalline axis vanishes when 
\beq k_3\propto \langle h_1(\bk)h_2(\bk)v_x(\bk)v_y(\bk)\rangle_{\mathrm{FS}}\label{k3latt}\eeq
does.  Here $\langle \cdots\rangle_{\mathrm{FS}}$ denotes an integral over the Fermi surface.

For a $d_{x^2-y^2}+id_{xy}$ order parameter on a square lattice, $h_1 = \cos k_x-\cos k_y$, $h_2 = \sin k_x\sin k_y$, and (\ref{k3latt}) vanishes by symmetry.  Turning to a triangular lattice, aligning one of the symmetry axes with the $x$-axis, $v_x=\partial_{k_x}\xi_{\bk}$ and $v_y=\partial_{k_y}\xi_{\bk}$ with $\xi_{\bk} = -2t[2\cos (\sqrt{3}k_y/2)\cos(k_x/2)+\cos k_x]$.   Using the same forms for the order parameters as we used in our numerical BdG calculations, we find that (\ref{k3latt}) vanishes for $f$-wave, but not chiral $p$- and $d$-wave, consistent with our numerical BdG results. 

Also consistent with our numerical results, the full GL coefficient (\ref{k3}) for chiral $d$-wave is much smaller than that for chiral $p$-wave, suggestive of a smaller current.  In GL, this suppression is due to the multiple sign changes of the $d$-wave order parameter around the Fermi surface, leading to a partial cancellation.  In the continuum limit, this partial cancellation becomes complete, tying into our continuum BdG results.  

To make contact with the total angular momentum discussion in Sec.~\ref{sec:Lztotal} and the ``Chern--Simons-like'' current (\ref{jCS}), we now discuss the modifications to GL for the case where there is a spatially varying $A_0(\br)$.  A relevant discussion can be found in Ref.~\onlinecite{Furusaki:01}. The disinterested reader may pass over this and proceed directly to the Discussion without losing continuity. 

The presence of a spatially varying potential $A_0(\br)$ leads to new gradient terms in the GL expansion of the form
\beq f_{\mathrm{GL}} = c^{\mu\nu}_{\alpha\beta}\left[\psi^*_{\alpha}(\partial_{\mu}\psi_{\beta})(\partial_{\nu}A_0)+\mathrm{c.c.}\right]+\cdots\label{fGLCS},\eeq
in addition to (\ref{eq:GL1}).  Here, $\mu,\nu$ denote Cartesian coordinates (e.g., $x$ and $y$) while $\alpha,\beta=1,2$ denote the components of the order parameter.  The real-valuedness of the free energy in conjunction with $U(1)$ gauge symmetry requires $c^{\mu\nu}_{\alpha\beta}\equiv c^{\mu\nu}\epsilon_{\alpha\beta}$, where $\epsilon_{\alpha\beta}$ is again the 2D Levi--Civita symbol.  The current arising from this is
\beq j_{\mu} =\frac{\partial f_{\mathrm{GL}}}{\partial (\partial_{\mu}\theta)} =  -2c^{\mu\nu}\Delta_{01}\Delta_{02}(\partial_{\nu} A_0).\label{CScurrent}\eeq

Equation (\ref{fGLCS}) leads to the following microscopic definition:
\beq c^{\mu\nu}\equiv \frac{1}{2\Delta_{01}\Delta_{02}}\lim_{\bq\to 0}\left.\frac{\partial \chi_{0\mu}(\bq)}{i\partial q_{\nu}}\right|_{\Delta_{01}=\Delta_{02}=0}.\label{cdef}\eeq
Here $\chi_{0\mu} \equiv (2\beta)^{-1}\sum_{\bk,\omega_n}v_{\mu}(\bk)\mathrm{tr}[\bG_0(\bk+\tfrac{\bq}{2},i\omega_n)\hat{\tau}_3\bG_0(\bk-\tfrac{\bq}{2},i\omega_n)]$ is the static current-charge correlator per spin, where $\bG_0(\bk,i\omega_n)$ is the appropriate matrix Nambu--Gorkov Green's function (as a function of the Matsubara frequency $\omega_n$) and $\hat{\tau}_3$ is the Pauli spin matrix.  This correlation function is readily evaluated at all temperatures:
\begin{align} \lim_{\bq\to \b0}\frac{\partial \chi_{0\mu}(\bq)}{i\partial q_{\nu}} =& \Delta_{01}\Delta_{02}\sum_{\bk}\frac{v_{\mu}(\bk)}{4E^3_{\bk}}\tanh(\beta E_{\bk}/2)\nonumber\\&\times \left[h_2(\partial_{k_{\nu}}h_1) - h_1(\partial_{k_{\nu}}h_2)\right],\label{currentcharge}\end{align}
where $E_{\bk} \equiv \sqrt{\xi^2_{\bk} + |\Delta_{\bk}|^2}$ is the bulk BCS quasiparticles dispersion.  

Using (\ref{Chern}), (\ref{currentcharge}), and $\partial\chi_{0\mu}/\partial q_{\nu} = -\partial \chi_{0\nu}/\partial q_{\mu}$, one sees that at $T=0$, modulo terms ${\cal{O}}(\Delta^2_0/E^2_F)$ that vanish in the weak-coupling limit, the Chern number is given by
\beq \frac{C}{8\pi} = \lim_{\bq\to \b0}\frac{\partial \chi_{0\mu}(\bq)}{2i\partial q_{\nu}} \epsilon_{\nu\mu}.\eeq
Combining this result with (\ref{CScurrent}) and (\ref{cdef}) gives the result (\ref{jCS}) for the $T=0$ current.    

At $T=T_c$, (\ref{cdef}) and (\ref{currentcharge}) give
\begin{align} c^{\mu\nu} =\sum_{\bk}\frac{v_{\mu}(\bk)}{4\xi^3_{\bk}}\tanh(\beta_c \xi_{\bk}/2)\left[h_2(\partial_{k_{\nu}}h_1) - h_1(\partial_{k_{\nu}}h_2)\right].\label{currentchargeGL}\end{align}
In conjunction with (\ref{CScurrent}), this shows that the ``Chern--Simons'' current (\ref{jCS}) at $T=0$ smoothly evolves into a contribution $\propto c^{\mu\nu}\Delta_{01}(T)\Delta_{02}(T)$ near $T_c$.  The momentum-space integrand involved with $c^{\mu\nu}$ has the same structure as that for the Chern number in the weak-coupling limit and as a result, $c^{\mu\nu}$ will not vanish as long as the Chern number does not. Moreover, in the soft edge limit, the two components of order parameter have the same spatial variation and the contribution to the current from (\ref{jGL}) vanishes. In this limit, the current is given by (\ref{CScurrent}) and does not vanish for any nonzero $m$. It is only in the sharp edge case, where $A_0=0$ in the superconductor, that (\ref{k3latt}) provides the condition for the edge current to vanish.

\section{Discussion}
\label{sec:discussion}

Using semiclassical Bogoliubov-de Gennes (BdG), we have shown that the edge current for any chiral superconductor other than $p$-wave vanishes exactly in the weak-coupling, continuum limit.  Using numerical BdG and Ginzburg--Landau (GL) calculations, this result was generalized  to a variety of lattice models.  
Specifically, we find nonzero integrated currents for $p_x+ip_y$ on square and triangular lattices, and $d_{x^2-y^2}+id_{xy}$ on a triangular lattice.  We find very small integrated currents (which vanish in the limit $\Delta_0/E_F\to 0$, neglecting the possible growth of sub-dominant order parameters near the surface) for $d_{x^2-y^2}+id_{xy}$ on a square lattice, and $f_{x(x^2-3y^2)}+if_{y(3x^2-y^2)}$ on a triangular lattice.  Noting that our zero-temperature BdG results are in complete agreement with GL on the matter of which systems we have studied exhibit edge currents, we expect that the vanishing of the Fermi surface integral (\ref{k3latt}) gives a simple condition for the edge current to vanish in both continuum and lattice systems.  Although we have not explored mixed states such as chiral $d_{xy}+is$ which are not eigenstates of $\hat{L}_z$, (\ref{k3latt}) also shows that this state will give rise to a nonvanishing edge current in the continuum, as expected from semiclassical BdG analyses~\cite{Rainer:98,Horovitz:03}.  

For the combinations of superconducting states and lattices that have been studied, the existence of an edge current for a particular state coincides with the order parameter components both transforming like basis functions of the same 2D irreducible representation of the lattice symmetry group.  On the square lattice, for instance, $p_x$ and $p_y$ form a basis for the 2D representation $E$, whereas  $d_{xy}$ and $d_{x^2-y^2}$ are bases for two different representations, $B_1$ and $B_2$.  Generally one would expect chiral states to be energetically favourable only when the two components are degenerate or nearly degenerate, and our calculations suggest they will generally have non-zero currents under such conditions, albeit reduced currents for angular momenta greater than 1.

In the remainder of this concluding section, we discuss possible implications of our results for some candidate chiral superconductors. 

\begin{center}
\begin{table}
{\renewcommand{\arraystretch}{1.3}
\begin{tabular}{|c|c|c|}
\hline
OP symmetry; lattice & Integrated current ?& Degenerate?\\ \hline 
$p$-wave; continuum & yes & yes \\ \hline
$d$-wave; continuum & no & yes \\ \hline
$p$-wave; square& yes & yes\\ \hline
$d$-wave; square& no& no\\ \hline
$p$-wave; triangle & yes & yes\\ \hline
$d$-wave; triangle &yes &yes\\ \hline
$f$-wave; triangle &no & no \\ \hline
\end{tabular}}
    \caption{\label{tab:table1}Order parameter (OP) and lattice symmetries and their relation to the existence of an integrated current.  By ``degenerate'', we mean that the two order parameter components transform with the same two-dimensional irreducible representation; details are given in the text.  For chiral states in the continuum, \emph{all} states with $m>1$ have vanishing edge currents.}
     \end{table}
\end{center}

After superfluid $^3$He-$A$, the most studied candidate chiral superconductor to date is unquestionably $\Sr$~\cite{Mackenzie:03,Maeno:12,Kallin:12}. Whilst $\mu$SR~\cite{Luke:98} and Kerr effect~\cite{Xia:06} measurements are strongly suggestive of spontaneous time-reversal symmetry-breaking below $T_c$, as noted in the Introduction, SQUID magnetometry measurements have not seen evidence for edge currents~\cite{Kirtley:07}.   Away from the clean-edge limit explored in the present paper, disorder~\cite{Ashby:09}, gap anisotropy~\cite{Huang:14}, and other edge effects~\cite{Sauls:11,Imai:12,Lederer:14,Bouhon:14} can have pronounced effects on the edge current, reducing them significantly.   Here we speculate on another possibility, that $\Sr$ is a chiral superconductor, but not $p$-wave.  We emphasize that while we know of no microscopic reason why e.g., chiral $f$-wave pairing should be favoured on a square lattice such as that for $\Sr$ (emphasizing that the order parameter components are not expected to be degenerate), this scenario would not necessarily be incompatible with the above experiments.  

There exist some early proposals for chiral $f$-wave states such as  $(k_x^2-k_y^2)(k_x+ik_y)$, $k_xk_y (k_x+ik_y)$, and $k_z^2(k_x+ik_y)$ in $\Sr$~\cite{Hasegawa:00,Graf:00,Won:00,Dahm:00}.  These correspond to $m=1$, however, and hence, are expected to give rise to substantial edge currents.  On the other hand, the 3D chiral $f_{z (x+iy)^2}$ state would exhibit the same (vanishing) edge current properties as a $d_{x^2-y^2}+id_{xy}$ state on a square lattice, although as noted before, the components are not expected to be degenerate on such a lattice.  

The vanishing of the edge current for such a state need not be
incompatible with $\mu$SR experiments, generally interpreted in terms
of spontaneous edge currents at domain walls separating regions of
opposite chirality~\cite{Luke:98}, as well as around impurities,
including the muons themselves.  The irregular structure of the domain
walls as well as the the local nature of perturbing impurities means
that some local currents would likely arise along irregular edges.  As much has been seen in BdG studies of chiral
$d+id$-wave~\cite{Graf:00A} and $d+is$~\cite{Lee:09} superconductors.
In Appendix~\ref{sec:GLextra}, we show how to extend the GL theory
presented here to describe edge currents along non-crystalline axes. For situations where the edge current vanishes along a crystalline axis, it does not vanish along other edges.

Another major piece of evidence in favour of time-reversal symmetry-breaking superconductivity in $\Sr$ is the appearance of a Kerr effect below $T_c$~\cite{Xia:06} (also seen in $\Up$~\cite{Schemm:14}).  In continuum systems, similar to our results for the edge current, this effect vanishes for all chiral states except for chiral $p$-wave~\cite{Goryo:08}.     Away from the continuum limit, however, an intrinsic Kerr effect arises from multiband transitions~\cite{Wysokinski:12,Taylor:12}.  Although we cannot make any definitive statement about whether multiband chiral $f$-wave superconductivity on a  square lattice would allow for a Kerr effect without a specific model,  we note that the Fermi surface integral (\ref{k3latt}) involved with the edge current is quite different than that involved in the intrinsic Kerr effect~\cite{Taylor:12}.

Some other candidate chiral superconductors that have recently attracted interest are $\Up$~\cite{Joynt:02}, Na$_x$CoO$_2 \cdot y$H$_2$O~\cite{Kiesel:13}, and SrPtAs\cite{Fischer:14}, all of which are conjectured to be either chiral $d$-wave or $f$-wave superconductors with an in-plane chiral $d$-wave component.  Without detailed knowledge about the structure of the order parameters, we again cannot draw any firm conclusions about the edge currents for these candidate gap symmetries.  Our results suggest that one would expect such states to exhibit edge currents, albeit reduced from that of chiral $p$-wave pairing.

{\textit{Note added---}}As this manuscript was being prepared for submission, a preprint~\cite{Tada:14} appeared which has some overlap.  Focussing on the problem of the total angular momentum in the continuum limit,  the authors of Ref.~\onlinecite{Tada:14} find that the total angular momentum vanishes to order $\Delta_0/E_F$ in the weak-coupling BCS limit for all states with $m>1$, consistent with our results. They also extend these results to the BEC limit of the crossover, where they derive the result given by (\ref{Lmexpected}) for all $m$.  These results have also been commented on by Volovik~\cite{Volovik:14}.
\acknowledgements
We thank Graeme Luke and John Berlinsky for helpful discussions. This work is supported by NSERC and CIFAR and by the Canada Research Chair and Canada Council Killam programs and the National Science Foundation under Grant No. NSF PHY11-25915 (CK). 

\appendix
\section{Dirac equation}
\label{sec:semiclassical}
In this section, we show how to map the semiclassical limit of the BdG equations (\ref{BdG}) and (\ref{BdGsolns}) onto the one-dimensional twisted-mass Dirac equation~\cite{Stone:04} and use its solution to derive (\ref{dispersion}) and (\ref{Q}).

Substituting (\ref{BdGsolns}) into (\ref{BdG}) and making the usual weak-coupling  and semiclassical approximations [$\mu=E_F, \partial^2_y a_{\sigma}(y)\ll k_F \partial_ya_{\sigma}(y),\partial^2_y b_{\sigma}(y)\ll k_F \partial_yb_{\sigma}(y)$],  the BdG equation reduces to the two one-dimensional Dirac equations 
\beq \left(\begin{array}{cc} -i\sigma \partial_x& \Delta_0e^{im\theta}\\ \Delta_0e^{-im\theta}&i\sigma \partial_x
\end{array}\right)\P_{\sigma} = E\P_{\sigma},\label{Dirac1}\eeq
where, as before, $\P^{\dagger}_{\sigma}\equiv [a_{\sigma},b_{\sigma}]$, $\sigma = \pm$, and we have defined 
\beq x\equiv y/\hbar v_F\sin\theta,\label{xdef}   \eeq
with  $v_F\equiv \hbar k_F/m^*$.  Taking the complex conjugate of the $\sigma=-$ Dirac 
equation,  these two equations can be combined into a single ``twisted mass'' Dirac equation,
\beq \left(\begin{array}{cc} -i \partial_x& \Delta_0e^{i\phi(x)}\\ \Delta_0e^{-i\phi(x)}&i \partial_x
\end{array}\right)\bar{\Psi}= E\bar{\Psi},\label{Dirac2}\eeq
for the composite spinor  $\bar{\Psi}\equiv \Theta(-x)\P_-(x) + \Theta(x)\P_+(x)$, where
\beq \phi(x) = -\Theta(-x)m\theta +\Theta(x)m\theta.\label{phitwist}\eeq

The two-dimensional edge problem has thus been mapped onto a one-dimensional problem where the phase of the order parameter is twisted across a domain at $x=0$ from $\phi_L=-m\theta$ on the left-hand side to $\phi_R=m\theta$ on the right.  The boundary condition $\P_{+}(y=0) = \P_-(y=0)$ in the original two-dimensional problem gets mapped onto the condition that $\bar{\Psi}(x)$ is continuous across $x=0$.  The integrated quasiparticle density $\sum_{\sigma}\P^{\dagger}_{\sigma}\P_{\sigma}$ needed to calculate the edge current is given 
by the ``charge''  $Q_m \equiv \sum_n\int^{\infty}_{-\infty} dx |\chi_n(x)|^2$ accumulated in the vicinity of the domain wall, where $\chi_n$ are the eigenstates of (\ref{Dirac2}) for a given magnetic quantum number $m$.  

The solution of (\ref{Dirac2}) is discussed at length in Ref.~\onlinecite{Stone:04}.  The only difference in our case is that the phase is twisted between $-m\theta$ and $m\theta$ instead of between $-\theta$ and $\theta$.  
This difference manifests itself in two ways.  First, everywhere in the appendix of Ref.~\onlinecite{Stone:04} where $\Phi\equiv \phi_L-\phi_R$ appears, we replace this with $-2m\theta$.  Second, for the calculation of the edge state properties, the mismatch between the $\sin\theta$ factor that arises when mapping back to the original $y$-coordinate [c.f. (\ref{xdef})] and the $\sin m\theta, \cos m\theta$ factors that arise in the solutions of (\ref{Dirac2}) and (\ref{phitwist})  leads to piecewise constraints when $m\neq 1$.   (\ref{Dirac2}), for instance,  supports a bound-state solution~\cite{Stone:04}
\beq \chi_0(x>0/x<0) \propto \left[\begin{array}{c} E^{(0)} \pm i\kappa + \Delta_0\\ E^{(0)} \mp i\kappa + \Delta_0\end{array}\right]e^{\mp\kappa x},\label{bssoln}\eeq
with $\kappa = \Delta_0 \sin m\theta$.  
Using (\ref{xdef}) and (\ref{bssoln}), boundedness in the original $y$-space means that $\kappa/\sin\theta = \Delta_0(\sin m\theta/\sin \theta)$ must be positive for all $\theta$.  This constraint $(\sin m\theta/\sin\theta>0)$ plus continuity $[\chi_0(0^+) = \chi_0(0^-)]$ leads to the result (\ref{dispersion}).  

Turning to the continuum bulk states, the charge $Q_m$ is calculated in exactly the same way as in Ref.~\onlinecite{Stone:04} with the replacement  $\Phi\equiv -2m\theta$ in e.g., their Eq.~(A13).  The same considerations that lead to Eq.~(A16) in Ref.~\onlinecite{Stone:04} yield (\ref{Q}).  

\section{Ginzburg--Landau theory for edges not aligned with the crystalline axes}
\label{sec:GLextra}

Here we generalize the GL expression (\ref{k3}) to allow for the possibility of currents along edges that are not parallel with crystalline axes.  Implicit in the appearance of $k_3$ in the GL free energy density (\ref{eq:GL1}) is that it describes the energy cost associated with a spontaneous current [$U(1)$ phase] along the $y$-axis and spatial modulation of the amplitude of the order parameter along $x$ (and vice-versa), as would happen if there was an edge parallel to the $y$-axis ($x$-axis).  One can generalize the definition of $k_3$ to allow for arbitrary orientation of the amplitude gradient, with the edge and resulting current perpendicular to this: $k_{3}(\phi) \equiv \partial^2\Gamma^{-1}_{12}(\bq,0)/\partial q'_x\partial q'_y$, where $\bq'\equiv [q'_x,q'_y]$ is rotated by $\phi$ with respect to $\bq$.
This leads to
\begin{align} k_3(\phi)& \equiv \sin\phi\cos\phi \left[\frac{\partial^2\Gamma^{-1}_{12}}{\partial q^2_x}-\frac{\partial^2\Gamma^{-1}_{12}}{\partial q^2_y}\right]\nonumber\\&+(\cos^2\phi-\sin^2\phi)\frac{\partial^2\Gamma^{-1}_{12}}{\partial q_x\partial q_y}.\label{k32}\end{align}
This describes the current along an edge oriented by an angle $\phi$ with respect to a crystalline axis.  

In the vicinity of an edge that is not parallel with a crystalline axis, we expect the order parameter to reorient itself to lower gradient energies, meaning that the $h_1$ and $h_2$ that enter this expression will be different.  For an edge not along an axis of symmetry of the crystal, an additional calculation would be required to compute the resulting order parameter.  Otherwise, symmetry and energetic arguments can be used to infer the correct form.   As an example, a  $\sin k_x + i\sin k_y$ order parameter on a cubic lattice will become $\sin k_x\cos k_y - \cos k_x\sin k_y + i(\sin k_x\cos k_y + \cos k_x\sin k_y)$ in the vicinity of the [11] edge; that is, it will simply be rotated in momentum space by $\pi/4$.   Likewise, assuming that the $d_{x^2-y^2} +id_{xy}$ order parameter on a cubic lattice is rotated by $\pi/4$ gives $h_1 = \sin k_x\sin k_y$ and $h_2 =  (\sin k_x\cos k_y)^2 - (\cos k_x\sin k_y)^2$.  The second line in (\ref{k32}) vanishes for $\phi=\pi/4$ while the first line involves a Fermi surface average of $h_1(\bk)h_2(\bk)(v^2_x-v^2_y)$, which also vanishes.  Thus, the generalized GL expression (\ref{k32}) predicts a vanishing edge current along the [11] edge as well as the [01] edge for a $d_{x^2-y^2}+id_{xy}$ order parameter on a square lattice.  We have also used (\ref{k32}) to confirm that $s+id_{x^2-y^2}$ on a square lattice supports a current along [11], even though there is none along [01]~\cite{Horovitz:03}.


\begin{thebibliography}{99} 

\bibitem{Ishikawa:77} M.~Ishikawa, Prog. Theor. Phys. \textbf{57}, 1836 (1977).
\bibitem{Mermin:80} N.~D. Mermin and P.~Muzikar, Phys. Rev. B \textbf{21}, 980 (1980).
\bibitem{Kita:98} T.~Kita, J. Phys. Soc. Jpn. \textbf{67}, 216 (1998).
\bibitem{Stone:04} M.~Stone and R.~Roy,  Phys. Rev. B {\bf 69}, 184511 (2004).
\bibitem{Sauls:11} J.~A. Sauls, Phys. Rev. B, \textbf{84}, 214509 (2011). 
\bibitem{Mackenzie:03} A.~P. Mackenzie and Y.~Maeno, Rev. Mod. Phys. \textbf{75}, 657 (2003).
\bibitem{Maeno:12} Y.~Maeno, S.~Kittaka, T.~Nomura, S.~Yonezawa, and K.~Ishida, J. Phys. Soc. Jpn. \textbf{81}, 011009 (2012). 
\bibitem{Kallin:12} C.~Kallin, Rep. Prog. Phys. \textbf{75}, 042501 (2012).
\bibitem{Kirtley:07} J.~R.~Kirtley, C.~Kallin, C.~W.~Hicks, E.-A.~Kim, Y.~Liu, K.~A.~Moler, Y.~Maeno, and K.~D.~Nelson, Phys. Rev. B \textbf{76}, 014526 (2007).  
\bibitem{Curran:14} P. J.~Curran, S. J.~Bending, W. M.~Desoky, A. S. Gibbs, S. L. Lee, and A. P. Mackenzie, Phys. Rev. B {\bf 89}, 144504 (2014).
\bibitem{Kallin:09} C.~Kallin and J.~Berlinsky, J.~Phys. Condens. Matter \textbf{21}, 164210 (2009).  
\bibitem{Ashby:09} P.~Ashby and C.~Kallin, Phys. Rev. B \textbf{79}, 224509 (2009).
\bibitem{Huang:14} W.~Huang, S.~Lederer, E.~Taylor, S.~Raghu, and C.~Kallin, \textit{in preparation}.
\bibitem{footnote} For a charged chiral superconductor, screening effects reduce the angular momentum by an amount $ \gtrsim \lambda/R$, where $\lambda$ is the penetration depth and $R$ is the radius of the disc.
\bibitem{Stone:08} M.~Stone and I.~Anduaga, Ann. Phys. \textbf{323}, 2 (2008). 
\bibitem{Matsumoto:95} M.~Matsumoto and H.~Shiba, J. Phys. Soc. Jpn. \textbf{64}, 4867 (1995). 
 \bibitem{Rainer:98} D.~Rainer, H.~Burkhardt, M.~Fogelstr\"{o}m, and J.~A.~Sauls, J. Phys. Chem. Solids. \textbf{59}, 2040 (1998). 
\bibitem{Horovitz:03} B.~Horovitz and A.~Golub, Phys. Rev. B \textbf{68}, 214503 (2003). 
\bibitem{Joynt:02} R.~Joynt and L.~Taillefer, Rev. Mod. Phys. \textbf{74}, 235 (2002). 
\bibitem{Schemm:14} E.~R.~Schemm, W.~J.~Gannon, C.~M.~Wishne, W.~P.~Halperin, and A.~Kapitulnik, Science \textbf{345}, 190 (2014).  
\bibitem{Volovik:88} G.E. Volovik, Sov. Phys. JETP {\bf 67}, 1804 (1988).
\bibitem{Goryo:98} J.~Goryo and K.~Ishikawa, Phys. Lett. A \textbf{246}, 549 (1998).
\bibitem{Volovik:99} G.~E.~Volovik, JETP Lett. \textbf{70}, 609 (1999).  
\bibitem{Leggett:80} A.~J.~Leggett,  in \textit{Modern Trends in the Theory of Condensed Matter}, edited by A. ~Pekalski and R.~Przystawa  (Springer--Verlag, Berlin, 1980).  
\bibitem{Jiang:08} Y.~Jiang, D-X.~Yao, E.~W.~Carlson, H.-D.~Chen and J.~P.~Hu, Phys. Rev. B {\bf 77}, 235420 (2008).
\bibitem{Nandkishore:12} R.~Nandkishore, L.~S.~Levitov, and A.~V.~Chubukov, Nature Phys. \textbf{8}, 158 (2012).  
\bibitem{Black:12} A.~M.~Black-Schaffer, Phys. Rev. Lett. \textbf{109}, 197001 (2012).
\bibitem{Braunecker:05} B.~Braunecker, P.~A.~Lee, and Z.~Wang, Phys. Rev. Lett. \textbf{95}, 017004 (2005). 
\bibitem{Kiesel:13} M.~Kiesel, C.~Platt, W.~Hanke, and R.~Thomale, Phys. Rev. Lett. \textbf{111}, 097001 (2013).
\bibitem{Fischer:14} M.~H. Fischer, T.~Neupert, C.~Platt, A.~P.~Schnyder, W.~Hanke, J.~Goryo, R.~Thomale, and M.~Sigrist, Phys. Rev. B \textbf{89}, 020509(R) (2014).
\bibitem{Sigrist:91} M.~Sigrist and K.~Ueda, Rev. Mod. Phys. \textbf{63}, 239 (1991).
\bibitem{Furusaki:01} A. Furusaki, M. Matsumoto and M. Sigrist, Phys. Rev. B {\bf 64}, 054514 (2001). 
\bibitem{note} Sauls [J.~A. Sauls, Adv. Phys. \textbf{43}, 113 (1994)] also finds that $k_3$ vanishes in the continuum for the chiral $k_z[(k^2_x-k^2_y) \pm 2ik_xk_y]$ state. 
\bibitem{Luke:98} G.~M.~Luke, Y.~Fudamoto, K.~M.~Kojima, M.~I.~Larkin, J.~Merrin, B.~Nachumi, Y.~J.~Uemura, Y.~Maeno, Z.~Q.~Mao, Y.~Mori, H.~Nakamura and M.~Sigrist, Nature \textbf{394}, 558 (1998). 
\bibitem{Xia:06} J.~Xia, Y.~Maeno, P.~T.~Beyersdorf, M.~M.~Fejer, and A.~Kapitulnik, Phys. Rev. Lett. \textbf{97}, 167002 (2006).
\bibitem{Imai:12} Y.~Imai, K.~Wakabayashi, and M.~Sigrist, Phys. Rev. B \textbf{85}, 174532 (2012).
\bibitem{Lederer:14} S.~Lederer, W.~Huang, E.~Taylor, S.~Raghu, and C.~Kallin, Phys. Rev. B \textbf{90}, 134521 (2014)
\bibitem{Bouhon:14} A.~Bouhon and M.~Sigrist, arXiv:1409.1516.  
\bibitem{Hasegawa:00} Y. Hasegawa, K. Machida and M. Ozaki, J. Phys. Soc. Jpn. {\bf 69}, 336 (2000).
\bibitem{Graf:00} M.J. Graf, A.V. Balatsky, Phys. Rev. B {\bf 62}, 9697 (2000).
\bibitem{Won:00} H. Won and K. Maki, Europhys. Lett. {\bf 52}, 427 (2000).
\bibitem{Dahm:00}  T.~Dahm, H.~Won, and K.~Maki, arXiv:cond-mat/0006301.
\bibitem{Graf:00A} M.~J.~Graf, A.~V.~Balatsky and J.~A.~Sauls, Phys. Rev. B \textbf{61}, 3255 (2000). 
\bibitem{Lee:09} W.-C.~Lee, S.-C.~Zhang and C.~Wu, Phys. Rev. Lett. \textbf{102}, 217002 (2009).
\bibitem{Goryo:08} J.~Goryo, Phys. Rev. B \textbf{78}, 060501(R) (2008).
\bibitem{Wysokinski:12} K.~I.~Wysoki\'nski, J.~F.~Annett, and B.~L.~Gy\"orffy, Phys. Rev. Lett. \textbf{108}, 077004 (2012).  
\bibitem{Taylor:12} E.~Taylor and C.~Kallin, Phys. Rev. Lett. \textbf{108}, 157001 (2012).
\bibitem{Tada:14} Y.~Tada, W.~Nie, M.~Oshikawa, arXiv:1409.7459.  
\bibitem{Volovik:14} G.~E.~Volovik, arXiv:1409.8638.  




\end{thebibliography}
\end{document}